\documentclass[aps,12pt]{revtex4} 
\usepackage[T2A]{fontenc}
\usepackage{graphicx}
\begin{document}

\def\mev{\hbox{\ MeV}}
\def\lsim{\mathrel{\rlap{
\lower4pt\hbox{\hskip-3pt$\sim$}}
    \raise1pt\hbox{$<$}}}     
\def\gsim{\mathrel{\rlap{
\lower4pt\hbox{\hskip-3pt$\sim$}}
    \raise1pt\hbox{$>$}}}     



\title{ Canonical Strangeness and Distillation Effects \\ in Hadron
  Production } 

\author{ V.D.~Toneev$^{1,2}$ and A.S.~Parvan$^{2,3}$}
\affiliation{$1$Gesellschaft f\"ur Schwerionenforschung mbH,
  64291 Darmstadt, Germany  }
\affiliation{$2$Joint Institute for Nuclear Research, 141980 Dubna, Russia}
\affiliation{$3$Institute of Applied Physics, Moldova Academy of
  Sciences, MD-2028 Kishineu, Moldova}


\begin{abstract}
Strangeness canonical ensemble for Maxwell-Boltzmann statistics is
reconsidered for excited nuclear systems with non-vanishing net
strangeness. A new recurrence relation method is applied to find
the partition function. The method is first generalized to the
case of quantum  strangeness canonical ensemble. Uncertainties in
calculation of the $K^+/\pi^+$ excitation function are discussed.
A new scenario based on the strangeness distillation effect is put
forward for a possible explanation of anomalous strangeness
production observed at the bombarding energy $\sim 30$ GeV. The
peaked maximum in the $K^+/\pi^+$ ratio is considered as a sign of
the critical end-point reached in evolution of  the system  rather
than a latent heat jump emerging from the onset of the first order
deconfinement phase transition.
\end{abstract}


\maketitle

\section{Introduction}
In recent years, statistical thermal models have widely been used for
 analyzing heavy ion collisions (see the review
article~\cite{BRMS03}). These models describe available phase space
at the final stage of nuclear collision when inelastic interactions
between hadrons cease, {\em i.e., at the
chemical freeze-out}.   The  grand canonical (GC) ensemble is defined
in the large volume limit, so the volume parameter $V$ is an
extensive coefficient in the expressions for such quantities as mean
particle numbers controlled by the chemical potentials. Application
of the GC ensemble also implies that the number of produced
particles carrying a conserved charge is sufficiently large. In this
description the net value of a given charge (e.g. electric charge,
baryon number, strangeness, charm {\em etc.}) fluctuates from event
to event. These multiplicity fluctuations  with respect to mean
 particle multiplicity can be neglected only if the particles carrying
the charges are abundant, {\em i.e.}, for the system of
a large volume and/or high temperature. Here the charge is
conserved on the average and, therefore, the GC description is
adequate.

In the opposite limit of low production rate, fluctuations of the
particle number can be as large as its event averaged value. In this
case, the charge conservation has to be implemented exactly in each
event. So the ensemble to be chosen is the canonical one. Then, the
$V$-independent fugacities  are replaced by nonlinearly
$V$-dependent canonical chemical factors. The exact conservation of
quantum numbers introduces certain constraints on the system
considered. In the formulation of thermodynamics of strongly
interacting matter one needs to implement these constraints  that
are originally governed by internal symmetry of the
Hamiltonian~\cite{RT80,HR85,RKT03}. Consequently, the equilibrium
distribution of particle multiplicities can differ from that
expected in the GC limit.

In principle, for an equilibrium  system all conserved charges
should be treated canonically. However, in a good approximation,
baryonic and electric charge conservation can be considered
grand-canonically even for high-energy $pp$ collisions. This
treatment results only in a slight overestimation of the canonical
effects~\cite{KB02-1}. On the other hand, the importance of exact
strangeness conservation for hadron production in heavy ion
collisions has been clearly demonstrated some time ago in
Refs.~\cite{CR99,HRT00-1}, where closed expressions for
suppression of chemical factors were derived in terms of classical
Maxwell-Boltzmann statistics for the global strangeness neutrality
condition $S=0$.

In this paper, we reconsider the derivation of the strangeness
constraint for nuclear systems with non-vanishing net strangeness
$S$ applying two different methods for evaluating the partition
function and ensemble averages: the direct method and  the method
of recurrence relations~\cite{R58}. In the line of
Refs.~\cite{PTP00,Par04}, the second method allows us first to
generalize our canonical strangeness evaluation to the case of
quantum statistics. Two Appendixes contain details on the
calculation of canonical quantities by recurrence relations. The
derived formulae will be used to analyze the kaon excitation
function from heavy ion collisions and particular attention will
be paid to the discussion of an anomalous maximum of excitation
functions at the beam energy near 30 GeV. A new possible scenario
of its formation as a manifestation of the critical end-point is
advanced.

\section{Classical statistics}

Let us construct the thermodynamic partition function  $Z_S$ with the
exact strangeness conservation for a relativistic perfect gas of $N$
hadron species at temperature $T$ in space volume $V$. Every hadron
species $\alpha$ ($\alpha=1,\ldots,N$) is characterized by the
baryonic $b_{\alpha}$, electric $q_{\alpha}$ and strangeness
$s_{\alpha}$ charges,  the mass $m_{\alpha}$ and the spin
degeneration factor $g_{\alpha}$. Let us assume that this particle
ensemble is described by the Maxwell-Boltzmann statistics. In the
occupation number $\{\nu_{\alpha p}\}$ representation, the
constraint for an exact conservation of the net strangeness $S$ can
be directly imposed on the grand-canonical partition function
conserving the average baryonic and electric charges~:
\begin{eqnarray}
\label{1a} Z_{S} &=& \sum\limits_{\{\nu_{\alpha p}\}}
\delta(\sum\limits_{\alpha,p} s_{\alpha}\nu_{\alpha p} - S) \
\frac{1}{\prod\limits_{\alpha,p}\nu_{\alpha p}!} \
e^{-\beta\sum_{\alpha,p}\nu_{\alpha p} \varepsilon_{\alpha p}} \ ,
\end{eqnarray}
with
\begin{eqnarray}\label{2a} \varepsilon_{\alpha p} &=& \sqrt{\vec
p^2+m_{\alpha}^2}-\mu_B b_{\alpha}-\mu_Q q_{\alpha}~.
\end{eqnarray}
Here $\beta=1/T$, $\mu_{B}, \mu_{Q}$ are  baryonic and electric
chemical potentials, respectively, and  multi-index $p\equiv (\vec
p, { \sigma},\ldots)$ includes particle momentum $\vec p$, spin
projection ${\sigma}$ and other quantum numbers. The summation in
(\ref{1a}) is carrying out over all permutations $\{\nu_{\alpha p}\}
$ of occupation numbers. In this representation, the ensemble
average for the operator $O(\{\nu_{\alpha p}\})$  is defined as
follows~:
\begin{eqnarray}\label{3a} \langle O\rangle_{S}  &=&
\frac{1}{Z_S}\sum\limits_{\{\nu_{\alpha p}\}}
\delta(\sum\limits_{\alpha,p} s_{\alpha} \nu_{\alpha p} - S) \
O(\{\nu_{\alpha p}\}) \ \frac{1}{\prod\limits_{\alpha,p} \
\nu_{\alpha p}!} \ e^{-\beta\sum_{\alpha,p} \nu_{\alpha p} \
\varepsilon_{\alpha p}}~.
\end{eqnarray}
These basic expressions will be used below for finding the  mean
occupation numbers and mean particle multiplicities.

\subsection{The direct method for solving the partition function}

The partition function $Z_S$  can be summed up over the momentum
$\vec p$ by inserting the identical unit in the form   $
\sum_{\{n_{\alpha}\}} \prod_{\alpha}\delta(\sum_{p}\nu_{\alpha
p}-n_{\alpha})=1$ into Eq.(\ref{1a}). Then, the partition function is
reduced to
\begin{eqnarray}\label{12a}
Z_{S}&=&\sum_{\{n_{\alpha}\}} \delta(\sum_{\alpha}
s_{\alpha}n_{\alpha}-S) \
\prod_{\alpha}\frac{(Z_{\alpha}^1)^{n_{\alpha}}}{n_{\alpha}!}~,
\end{eqnarray}
where the grand-canonical single particle partition function
\begin{equation}\label{10a}
  Z_{\alpha}^1 = \sum_{p}
e^{-\beta\varepsilon_{\alpha p}}=e^{\beta (\mu_B \ b_{\alpha}+\mu_Q
\ q_{\alpha})} \ \frac{g_{\alpha}V}{(2\pi)^3} \int d^3\vec p \
e^{-\beta \sqrt{\vec p^2 +m_{\alpha}^2}}~.
\end{equation}
 Using the integral representation for the Kronecker symbol in
 Eq.(\ref{12a}), one can carry out summation over all particle numbers
$\{n_{\alpha}\}$  by means of  series expansion of the exponent.
 So we have
\begin{equation}\label{13a}
Z_{S} = \frac{1}{2\pi} \int\limits_{0}^{2\pi} d\phi \
\exp\left[-\imath S\phi + \sum\limits_{\alpha}Z_{\alpha}^{1}
e^{\imath s_{\alpha}\phi}\right]~.
\end{equation}
One should note that this representation of $Z_S$ based on the
classical statistical 
partition function (\ref{1a})  coincides exactly with the starting
expression used in~\cite{BRMS03,CR99,HRT00-1}.

Let us proceed now  from the summation over species index $\alpha$ to
the sum over all available strangeness values $s$ for individual
particles in the system.  So we substitute into Eq.(\ref{13a}) the
identical unit in the form $\sum_{s=-s_{max}}^{s_{max}}
\delta_{s,s_{\alpha}} = 1$, which is valid for any species $\alpha$
under condition $|s_{\alpha}|\le s_{max}$
\begin{equation}\label{17a}
Z_{S} = \frac{1}{2\pi} \int\limits_{0}^{2\pi} d\phi \
\exp\left[-\imath S\phi+
\sum\limits_{s=-s_{max}}^{s_{max}}S_{s}e^{\imath s\phi}\right]~,
\end{equation}
where
\begin{eqnarray}
\label{16a} S_{s} &=&
\sum_{\alpha}\delta_{s,s_{\alpha}}Z_{\alpha}^{1}~,
\;\;\;\;\;\;\;\;\; -s_{max}\leq s \leq s_{max}~.
\end{eqnarray}
By rewriting  the exponent of the sum over $-s_{max}\le s\le
s_{max}$  in Eq.(\ref{17a}) as the product of exponents and
expanding every exponent into series, $e^x=\sum x^n/n!$, we get
\begin{equation}\label{14a}
\exp\left[\sum\limits_{s}S_{s}e^{\imath s\phi}\right] =
\sum\limits_{\{N_s\}} \exp\left[\sum\limits_{s}\imath s
N_s\phi\right] \ \prod\limits_{s} \frac{S_{s}^{N_s}}{N_s!}~,
\end{equation}
where  $N_s$ are non-negative integer numbers. Here the following
short-hand notation has been used for the sum $\sum_{\{N_s\}} \equiv
\sum_{N_{-s_{max}}}\ldots \sum_{N_{s_{max}}}$. After substitution of
Eq.(\ref{14a}) into (\ref{17a}) and integration with respect to
$\phi$,  the strangeness canonical partition function $Z_{S}$ is
reduced to the following equation~:
\begin{eqnarray}\label{15a}Z_{S}&=&\sum_{\{N_{s}\}}
\delta(\sum_{s=-s_{max}}^{s_{max}} s N_{s}-S) \
\prod_{s=-s_{max}}^{s_{max}}\frac{S_{s}^{N_{s}}}{N_{s}!}~.
\end{eqnarray}

If the integral representation  for the Kronecker symbol is used in
Eq.(\ref{15a}), we come back to Eq.(\ref{17a}) for the partition
function in the statistical model~\cite{BRMS03}.

A number of sums over the set $\{N_s\}$ in Eq.(\ref{15a}) can be
reduced. For this one should introduce the identity
\begin{equation}\label{18a}
\prod\limits_{s=1}^{s_{max}} \sum\limits_{n_{s}=-\infty}^{\infty}
\delta(N_{s}-N_{-s} - n_{s}) =1~
\end{equation}
into (\ref{15a}), to be valid for every fixed number set $\{N_{s}\}$, 
and change the order of summing with respect to $\{n_{s}\}$ and
$\{N_{s}\}$. Then,  using the Kronecker symbol in Eq.(\ref{18a}),
the summation for the set $\{N_{s}\}$ can be done explicitly. It is
noticed that these sums are grouped by pairs for every $\mp |s|$ and
every pair results in the modified Bessel functions. Finally, for
Eq.(\ref{15a}) we get
\begin{equation}\label{19a}
Z_{S}=e^{S_0} \sum\limits_{n_1=-\infty}^{\infty}\ldots
\sum\limits_{n_{s_{max}}=-\infty}^{\infty}
\delta(\sum\limits_{s=1}^{s_{max}} s n_{s} - S) \
\prod\limits_{s=1}^{s_{max}} a_{s}^{n_{s}} \ I_{-n_{s}}(x_{s})~,
\end{equation}
where $I_n$ is the modified Bessel function and variables
\begin{equation}\label{20a}
x_{s} = 2\sqrt{S_{-s} \ S_{s}}~,  \;\;\;\; a_{s} =
\sqrt{S_{s}/S_{-s}}~, \;\;\;\;\;\;\;\;\; s=1,\ldots,s_{max}~.
\end{equation}
Taking into account Eq.(\ref{16a}) we see that $x_s \sim V$ while $a_s$
are independent of the system volume. In specific calculations the
maximal hadron strangeness $s_{max}$ is not large, which simplifies
essentially Eq.(\ref{19a}), since the number of infinite sums is
$(s_{max}-1)$. In particular, for $s_{max}=1,2,3$ the partition
function  (\ref{19a}) can be expressed as follows~:
\begin{eqnarray}
\label{21a} Z_{S} &=& e^{S_0} \sum\limits_{n=-\infty}^{\infty}
\sum\limits_{m=-\infty}^{\infty} a_1^{S-2n-3m} \ a_2^{n} \ a_3^{m} \
I_{S-2n-3m}(x_1) \ I_{n}(x_2) \ I_{m}(x_3)~, \;\;\;\;\;\;\;\;\; s_{max}=3~, \\
\label{22a} Z_{S} &=& e^{S_0} \sum\limits_{n=-\infty}^{\infty}
a_1^{S-2n} \ a_2^{n} \
I_{S-2n}(x_1) \ I_{n}(x_2)~, \;\;\;\;\;\;\;\;\; s_{max}=2~, \\
\label{23a} Z_{S} &=& e^{S_0} \ a_1^{S} \ I_{S}(x_1)~,
\;\;\;\;\;\;\;\;\; s_{max}=1~,
\end{eqnarray}
 where the symmetry property of the modified Bessel
function $I_{-n}=I_{n}$ has been used. Associative production of
multi-strange baryons is neglected in Eq.(\ref{23a}).
 This simplified equation  for $S\neq 0$ in the case of $s_{max}=1$ has been
 derived earlier in~\cite{BRMS03}. The result for $s_{max}=4$ can be found
in~\cite{Becat}. The case of full canonical treatment of both baryon
and strangeness $S\ne 0$ in a little bit different technique was given
in~\cite{CSW92}. However,   the electric charge conservation is taken
into account by neither canonically, nor grand canonically. It
simplifies calculations but   the final result differs from our
Eqs.(\ref{21a})-(\ref{23a}). The single particle partition function
$Z_{\alpha}^{1}$ given by Eq.(\ref{10a}) can be integrated resulting
in the well-known expression
\begin{equation}\label{24a}
Z_{\alpha}^{1}= \frac{g_{\alpha}V}{2\pi^2}
\frac{m_{\alpha}^2}{\beta} \ K_2(\beta m_{\alpha}) \ e^{\beta (\mu_B
b_{\alpha}+\mu_Q q_{\alpha})}
\end{equation}
with the Bessel function $K_2$.

\subsection{Method of recurrence relations}

Let us subdivide the set  $M$ of particles of all species into three
sub-sets $M_{\pm}$ and $M_{0}$ according to the value of individual
strangeness charge $s_{\alpha}$ of involved hadrons. For a fixed set
of occupation numbers $\{\nu_{\alpha p}\}$ we introduce notation of
positive $S_{+}$ and negative $S_{-}$ components of the net
strangeness $S$ of the system
\begin{eqnarray}
\label{4a1} \sum\limits_{\alpha\in M_{\mp}}\sum_{p} |\ s_{\alpha}| \
\nu_{\alpha p} &=& S_{\mp}~,
\end{eqnarray}
where  $M_+$ and  $M_-$ are the sets of all species  $\alpha$ for
which $s_{\alpha}>0$ and  $s_{\alpha}<0$, respectively. For the zero
component the following equality is fulfilled $\sum_{\alpha\in
M_{0}}\sum_{p} |s_{\alpha}| \nu_{\alpha p} =0$, since $M_0$ is the
set of all non-strange hadrons, $s_{\alpha}=0$. Let us substitute
into Eq.(\ref{1a}) the product of two identical unit operators
corresponding to a fixed set of occupation numbers $\{\nu_{\alpha
p}\}$~:
\begin{eqnarray}\label{4a}
\sum\limits_{S_{\mp}=0}^{\infty}\delta(\sum\limits_{\alpha\in
M_{\mp}}\sum_{p} |s_{\alpha}| \nu_{\alpha p}-S_{\mp}) = 1~.
\end{eqnarray}
Then for the strangeness canonical partition function we have
\begin{eqnarray}\label{5a}
Z_{S}&=& Z^{(0)}  \sum_{S_{+}=0}^{\infty}\sum_{S_{-}=0}^{\infty}
\delta(-S_{-}+S_{+}-S) \ Z_{S_{+}}^{(+)} \ Z_{S_{-}}^{(-)}~,
\end{eqnarray}
where the partition functions for positive $Z_{S_{+}}^{(+)}$ and
negative $Z_{S_{-}}^{(-)}$  components of the net strangeness $S$
have the following form in the case of classical statistics~:
\begin{eqnarray}\label{6a}
Z_{S_{\mp}}^{(\mp)} &=& \sum\limits_{\{\nu_{\alpha p}\}_{\alpha\in
M_{\mp}}} \delta(\sum\limits_{\alpha\in M_{\mp}}\sum_{p}
|s_{\alpha}| \ \nu_{\alpha p} - S_{\mp}) \
\frac{1}{\prod\limits_{\alpha\in M_{\mp}}\prod\limits_{p}\nu_{\alpha
p}!} \ e^{-\beta\sum_{\alpha\in M_{\mp}}\sum_{p} \ \nu_{\alpha
p} \ \varepsilon_{\alpha p}}~.
\end{eqnarray}
The positive $S_{+}$ and negative $S_{-}$  strangeness components
take integer non-negative values.  The appropriating partition
functions can be represented as follows (see Appendix A)~:
\begin{eqnarray}
  \label{8a}
  Z_{S_{\mp}}^{(\mp)} = \frac{1}{S_{\mp}}\sum_{\alpha\in M_{\mp}}
  |s_{\alpha}| Z_{\alpha}^1 \ Z_{S_{\mp}-|s_{\alpha}|}^{(\mp)} =
 \frac{1}{S_{\mp}} \sum_{s=1}^{s_{max}} s S_{\mp s} \
 Z_{S_{\mp}-s}^{(\mp)}~,
\end{eqnarray}
where the second equality in (\ref{8a}) was obtained by introducing
once more identical unit in the form $\sum_{s=1}^{s_{max}}
\delta_{s,|s_{\alpha}|} = 1$, to be valid for any species $\alpha$
under condition $|s_{\alpha}|\le s_{max}$. Here by definition
$Z_{0}^{(\mp)}=1$, $s_{max}=\max\{|s_{\alpha}|\}$ is maximal modulus
value of particle strangeness in the given system and auxiliary
functions are defined as
\begin{eqnarray}\label{9a}
    S_{\mp s} = \sum_{\alpha\in M_{\mp}}\delta_{s,|s_{\alpha}|} \
    Z_{\alpha}^1~, \;\;\;\;\;\;\;\;\; 1 \leq s \leq s_{max}~.
\end{eqnarray}
It is of great interest that mathematical structure of
Eqs.(\ref{8a}) allows one an exact recursive solution because the
partition function of the given strangeness component
$Z_{S_{\mp}}^{(\mp)}$ is expressed through the sum of partition
functions but with smaller index of the strangeness components.
This equation can be solved iteratively starting from the lowest
component $Z_{0}^{(\mp)}=1$. For calculating $Z_{S_{\mp}}^{(\mp)}$
this  procedure converges quickly as is seen from
Eq.(\ref{6a}): High values of $S_{\mp}$ in the argument of the
delta function should be compensated by large occupation numbers
which are exponentially suppressed with increasing negative
exponent. This situation reminds very much of finding the exactly
solvable partition function in the nuclear multifragmentation
model~\cite{PTP00,Par04,Mek} and in the context of quark-gluon
plasma~\cite{PR03} where the method of recurrence relations turned out to be
very powerful.

The partition function $Z^{(0)}$ with zero component of net
strangeness $S$ is reduced to the following generating function
$Z^{(0)}=e^{S_{0}}$ with $S_0 = \sum_{\alpha\in
M_{0}}Z_{\alpha}^{1}$, according to Eq.(\ref{9a}).

\subsection{Ensemble averages}

The mean occupation numbers can be found as ensemble averages
(\ref{3a}) for the operator  $O(\{\nu_{\alpha p}\})=\nu_{\alpha p}$.
Using mathematical technique  for summing over occupation numbers as
developed in~\cite{PTP00,Par04}, we get for {\em the mean occupation
numbers} the following result (see Appendix A)~:

\begin{equation}\label{25a}
\langle \nu_{\alpha p} \rangle_{S} = e^{-\beta \varepsilon_{\alpha
p}} \ \frac{Z_{S-s_{\alpha}}}{Z_{S}}~.
\end{equation}
{\em The mean particle number} of species $\alpha$ is obtained by
summing the mean occupation numbers  (\ref{25a}) over momentum $\vec
p$
\begin{equation}\label{26a}
\langle n_{\alpha} \rangle_{S} = \sum_{p}\langle \nu_{\alpha
p}\rangle_{S} = Z_{\alpha}^{1} \ \frac{Z_{S-s_{\alpha}}}{Z_{S}}~.
\end{equation}
Eq.(\ref{26a}) was given also in~\cite{Becat}. It is easy to
convince ourselves that the mean  number of particles with strangeness $s$ in
the system with the net strangeness $S$ can be represented as follow~:
\begin{equation}\label{27a}
\langle N_s \rangle_{S} = \sum_{\alpha} \delta_{s,s_{\alpha}} \
\langle n_{\alpha} \rangle_{S}= S_s \ \frac{Z_{S-s}}{Z_{S}}
\end{equation}
with $S_s $ defined by Eq.(\ref{9a}). As a particular case, for the
particle density in strangeness canonical ensemble $n_{\alpha} =
\langle n_{\alpha} \rangle_{S}/V$ we have:

\begin{eqnarray}
\label{28a} n_{\alpha} &=& \frac{Z_{\alpha}^{1}}{V}
\frac{e^{S_0}}{Z_{S}} \sum\limits_{n=-\infty}^{\infty}
\sum\limits_{m=-\infty}^{\infty} a_1^{S-s_{\alpha}-2n-3m} a_2^{n}
a_3^{m} \
I_{S-s_{\alpha}-2n-3m}(x_1)\ I_{n}(x_2)\ I_{m}(x_3), s_{max}=3, \\
\label{29a} n_{\alpha} &=& \frac{Z_{\alpha}^{1}}{V}
\frac{e^{S_0}}{Z_{S}} \sum\limits_{n=-\infty}^{\infty}
a_1^{S-s_{\alpha}-2n} \ a_2^{n} \
I_{S-s_{\alpha}-2n}(x_1) \ I_{n}(x_2)~, \;\;\;\;\;\;\;\;\; s_{max}=2~, \\
\label{30a} n_{\alpha} &=& \frac{Z_{\alpha}^{1}}{V} \
\left(\frac{S_{-1}}{\sqrt{S_{-1}S_1}} \right)^{s_{\alpha}} \
\frac{I_{S-s_{\alpha}}(x_1)}{I_{S}(x_1)}~, \;\;\;\;\;\;\;\;\;
s_{max}=1~.
\end{eqnarray}
Similarly to Eqs.(\ref{21a})-(\ref{23a}) only Eq.(\ref{28a}) takes
into account that strangeness of, for example,  kaons with
$s_{\alpha}=1$ can be compensated associatively by production of
any multi-strange hyperon. For non-strange hadrons,
$s_{\alpha}=0$, the particle density is
\begin{equation}\label{31a}
n_{\alpha} = \frac{Z_{\alpha}^{1}}{V}.
\end{equation}
as follows from  Eq.(\ref{26a}). For vanishing net strangeness,
$S=0$, Eqs.(\ref{28a})--(\ref{30a}) coincide with those in
Ref.~\cite{BRMS03} (see also references therein).

\section{Quantum statistics}
As compared to Eq.(\ref{1a}), in quantum statistics the partition
function with exact conservation of net strangeness and with in
average conservation of baryonic and electric charges can
represented as follows~:

\begin{eqnarray}\label{1b}
Z_{S} &=& \sum\limits_{\{\nu_{\alpha p}\}}
\delta(\sum\limits_{\alpha,p} s_{\alpha}\nu_{\alpha p} - S) \
e^{-\beta\sum_{\alpha,p}\nu_{\alpha p} \ \varepsilon_{\alpha p}}~,
\end{eqnarray}
where  $\varepsilon_{\alpha p}$ is defined by Eq.(\ref{2a}) and
occupation numbers  take values $\nu_{\alpha
p}=0,1,\ldots,K_{\alpha}$ with $K_{\alpha}=1$ for Fermi-Dirac
statistics and $K_{\alpha}=\infty$ for Bose-Einstein statistics. By means
of Eq.(\ref{4a}) the partition function (\ref{1b}) is reduced to
Eq.(\ref{5a}) with partition functions for positive and negative
strangeness components given by equation
\begin{eqnarray}\label{1bb}
Z_{S_{\mp}}^{(\mp)} &=& \sum\limits_{\{\nu_{\alpha p}\}_{\alpha\in
M_{\mp}}} \delta(\sum\limits_{\alpha\in M_{\mp}}\sum_{p}
|s_{\alpha}| \nu_{\alpha p} - S_{\mp}) \ e^{-\beta\sum_{\alpha\in
M_{\mp}}\sum_{p}\nu_{\alpha p} \ \varepsilon_{\alpha p}}~.
\end{eqnarray}
As is shown in  Appendix B,   the following recurrence relations are
fulfilled for these partition functions~:
\begin{eqnarray}
  \label{3b} Z_{S_{\mp}}^{(\mp)} = \frac{1}{S_{\mp}}\sum_{\alpha\in M_{\mp}}
  \sum_{l=1}^{[S_{\mp}/|s_{\alpha}|]}
  |s_{\alpha}| \ Z_{\alpha}^{l} \ Z_{S_{\mp}-l|s_{\alpha}|}^{(\mp)} =
 \frac{1}{S_{\mp}} \sum_{s=1}^{s_{max}}  \sum_{l=1}^{[S_{\mp}/s]} s \
S_{\mp s,l}
 \ Z_{S_{\mp}-l s}^{(\mp)}~,
\end{eqnarray}
where  $ Z_{0}^{(\mp)}=1$ and new auxiliary functions
\begin{eqnarray}\label{4b}
    S_{\mp s,l}&=& \sum_{\alpha\in M_{\mp}}
    \delta_{s,|s_{\alpha}|} \ Z_{\alpha}^{l}, \\
\label{5b}
    Z_{\alpha}^{l}&=& y_{\alpha l} \sum\limits_{p} e^{-\beta l
\varepsilon_{\alpha
    p}}~.
\end{eqnarray}
The quantity $y_{\alpha l}=(-1)^{l+1}$ for Fermi-Dirac statistics
and $y_{\alpha l}=1$ for Bose-Einstein statistics. Note that in the case of
Maxwell-Boltzmann statistics  $y_{\alpha l}=\delta_{l,1}$.
Eq.(\ref{3b}) exactly follows the structure of the corresponding
Eq.(\ref{8a}) for classical statistics and can be solved by the
recurrence relation method.

The partition function for zero strangeness component looks like
(see Appendix B)
\begin{equation}\label{6b}
    Z^{(0)}=\exp ({\sum_{l=1}^{\infty} S_{0,l}})~,
\end{equation}
with $S_{0,l}=\sum_{\alpha\in M_0} l^{-1} \ Z_{\alpha}^{l}$. Notice
that  under condition $|e^{-\beta  \varepsilon_{\alpha p}}|< 1$ the
partition function  $Z^{(0)}$ can be represented in more familiar
form~:
\begin{equation}\label{6b1}
  Z^{(0)}=\prod_{\alpha\in M_{0}}\prod_{p} (1\pm e^{-\beta
  \varepsilon_{\alpha p}})^{\pm 1}~,
\end{equation}
where the upper sign corresponds to Fermi-Dirac statistics and the
lower sign does to Bose-Einstein one.

As shown in Appendix B, the mean occupation numbers for hadrons with
negative and positive strangeness, $\alpha\in M_{\mp}$, are given as
\begin{eqnarray}\label{7b}
\langle \nu_{\alpha p}\rangle_{S}&=& \frac{Z^{(0)}}{Z_{S}}
\sum_{S_{+}=0}^{\infty}\sum_{S_{-}=0}^{\infty}
\delta(-S_{-}+S_{+}-S) \ Z_{S_{\pm}}^{(\pm)}
\sum\limits_{l=1}^{[S_{\mp}/|s_{\alpha}|]} y_{\alpha l} \
 e^{-\beta l
 \varepsilon_{\alpha p}} Z_{S_{\mp}-l |s_{\alpha}|}^{(\mp)},
  \alpha \in M_{\mp},
\end{eqnarray}
while those for non-strange particles are
\begin{eqnarray} \label{8b} \langle \nu_{\alpha p}\rangle_{S}&=&
\sum_{l=1}^{\infty} y_{\alpha l} \ e^{-\beta l \varepsilon_{\alpha
p}}~, \;\;\;\;\;\;\;\alpha \in M_{0}~. \ 
\end{eqnarray}
Under the same condition $|e^{-\beta  \varepsilon_{\alpha p}}|< 1$
we arrive at the familiar expression for mean occupation numbers
(\ref{8b})~:
\begin{equation}\label{8b1}
\langle \nu_{\alpha p}\rangle_{S}= \frac{1}{e^{\beta
\varepsilon_{\alpha p}}\pm 1}~, \;\;\;\;\;\;\;\alpha \in M_{0}~.
\end{equation}

To get the average number of particles $\alpha$ one should sum the
mean occupation numbers (\ref{7b}) and (\ref{8b}) over momentum
$p$. For strange particles  $\alpha \in M_{\mp}$ we have
\begin{eqnarray}\label{9b}
\langle n_{\alpha }\rangle_{S}&=& \frac{Z^{(0)}}{Z_{S}}
\sum_{S_{+}=0}^{\infty}\sum_{S_{-}=0}^{\infty}
\delta(-S_{-}+S_{+}-S) \ Z_{S_{\pm}}^{(\pm)}
\sum\limits_{l=1}^{[S_{\mp}/|s_{\alpha}|]}  Z_{\alpha}^{l} \
Z_{S_{\mp}-l |s_{\alpha}|}^{(\mp)}~, \;\;\;\;\;\;\;  \alpha \in
M_{\mp}~.
\end{eqnarray}
For non-strange hadrons $\alpha \in M_{0}$ the following relation is
valid~:
\begin{equation}\label{10b}
\langle n_{\alpha}\rangle_{S}= \sum_{l=1}^{\infty} Z_{\alpha}^{l}~,
\;\;\;\;\;\;\; \alpha \in M_{0}~.
\end{equation}
After integration  the single particle partition function (\ref{5b})
is reduced to ({\em cf.} with Eq.(\ref{24a}) )
\begin{equation}\label{11b}
Z_{\alpha}^{l}= y_{\alpha l} \ \frac{g_{\alpha}V}{2\pi^2} \
\frac{m_{\alpha}^2}{\beta l} \ K_2(\beta l m_{\alpha}) \ e^{\beta l
(\mu_B b_{\alpha}+\mu_Q q_{\alpha})}~.
\end{equation}
One should note that all formulae for Maxwell-Boltzmann statistics
from the proceeding section are recovered  by the simple
substitution $y_{\alpha l}=\delta_{l,1}$ into appropriate quantum
relations.

\section{Strangeness excitation function}
To reveal canonical effects in the obtained statistical relations, let
us consider  excitation functions for strange particle
multiplicities. These multiplicities are calculated along the
"experimental" freeze-out curve defined at any bombarding energy by
the temperature $T$ and the baryon chemical potential $\mu_B$~:
$$
\mu_{B}[in \ MeV] = \frac{1270}{1+\sqrt{s}/4.3},  
 \mbox{\hspace{10mm}}
T [in \ MeV]=\frac{170}{1+e^{-0.48(\sqrt{s}-3.8)}},
\;\;\;\;[\sqrt{s} \ is \ in \ GeV], 
$$

\noindent as
approximated in~\cite{BRMS03,Andronic}. The perfect hadronic gas
includes all mesons with  $m_\alpha \lsim 1.6$ GeV and baryons with
$m_\alpha \lsim 2$ GeV. In calculations of a particular hadron yield
the contribution from decays of heavier hadrons is taken into
account in addition to the primary yield
 $$  \langle n_{\alpha}\rangle =  \langle
  n_{\alpha}\rangle^{primary}+ \sum_{\gamma} Br(\gamma\rightarrow\alpha)
  \langle n_{\gamma}\rangle,$$
where the branching ratios $ Br(\gamma\rightarrow\alpha)$ are
taken from the Review of Particle Physics \cite{Hagiwara}.
Certainly the final answer will depend on whether only strong
decays or also weak ones are considered. The latter circumstance
is controlled by the experimental acceptance~\cite{Andronic}.

 The $K^+/\pi^+$ excitation function   is shown  in Fig.1 for systems
 formed in high-energy heavy-ion collisions. 
 The curve for the strangeness neutral system described by the
 classical Maxwell-Boltzmann statistics and without taking into  account
 weak decays   reproduces  closely the earlier statistical model results
(see~\cite{CR99,HRT00-1}
 as well as the review~\cite{BRMS03}). Note that two
 curves calculated within the classical ensemble by two above-discussed
 methods coincide exactly and are indistinguishable
 in the figure. Calculations have been carried out using a complete
 equation (\ref{28a}). If $s_{max}=1$  is assumed ({\em i.e.},
Eq.(\ref{30a}) is  applied), which physically implies that associative
 production of the $K^+$ mesons is accompanied by creation of
 single-strange-charge particles only, the  $K^+/\pi^+$ ratio
 decreases by about 5 $\%$ at the colliding energy $\sqrt s \sim 8$ GeV.
\begin{figure}[ht]
\includegraphics[height=10cm,width=10cm]{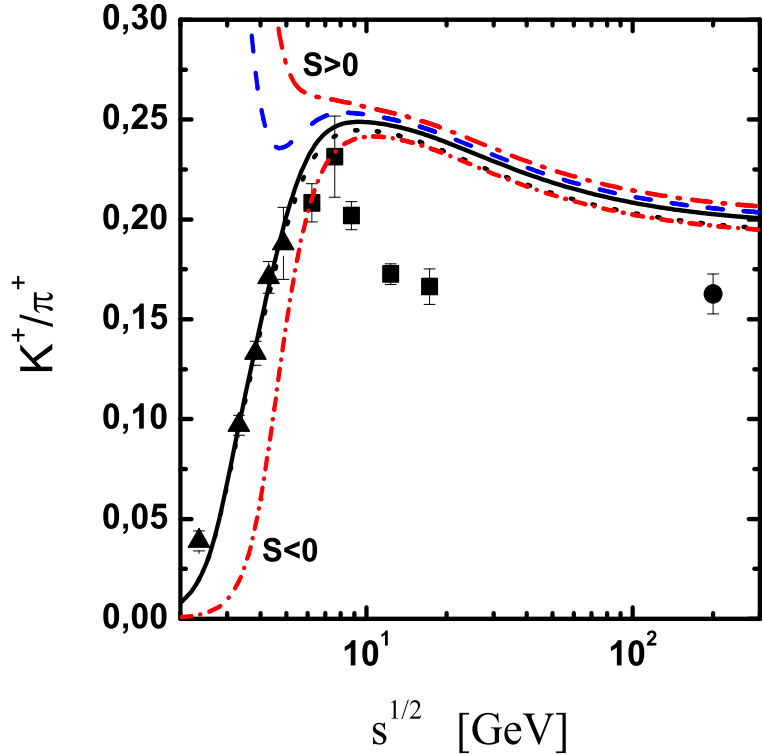}
\caption{Excitation function for relative strange particle
  production, $K^{+}/\pi^{+}$, in heavy ion  collisions. The symbols are
  experimental $4\pi$ data of E866 (triangles), NA49 (squares) and
  BRAHMS (circles) collaborations~\cite{BGKMS04,Brat1,B-NA49}.  The lines
are thermal model calculations for two types of statistics:
classical (solid) and quantum (dotted) for the net strangeness of
the system $S=0$ and classical Maxwell-Boltzmann ensemble for systems
  with non-vanishing net strangeness $S=$ 5, 10  (long dashed, dash
  dotted) and $S=$ -10 (short dash
  dotted), respectively.  The radius of the strangeness correlated sphere
is $R=7$ fm and electrical chemical potential $\mu_{Q}=0$ and
$s_{max}=3$. The additional contribution into $K^{+}$ and $\pi^{+}$
  from strong  decays of hadrons has been taken into account.
}\label{1}
\end{figure}

 The inclusion of quantum statistics slightly decreases the
 $K^+/\pi^+$ ratio  at $\sqrt s >6$ GeV.
 As is found, the contribution of weak
 decays of hadrons with the life-time $\tau < 10^{-10}$ sec  decreases
 this ratio by (15)$\%$  without changing the shape of the excitation
 function in agreement with recent finding  in~\cite{Andronic}.

It is noteworthy that due to local strangeness conservation,
 two volumes are usually introduced. The first $V$ is coming as a
normalization factor of the partition function (see, for example,
Eq.(\ref{31a}), which is dropped out if the particle ratio is
considered. The second volume, $V_c$, enters the argument $x$ of
the modified Bessel functions $I_n(x)$ in
Eqs.(\ref{28a})-(\ref{30a}) and characterizes the range of
strangeness correlations. It was shown~\cite{BRMS03} that for heavy
ion collisions $V_c$ is proportional to a number of participant
nucleons. The presented results correspond to a constant strangeness
correlation volume $V_c$ treated as a free parameters. Following
Ref.\cite{TR02}, this volume is taken as a sphere with the radius of
about $R\sim 7$ fm, which ensures the agreement with experiment for
the fast growth of the $K^+/\pi^+$ ratio at moderate energies.
Asymptotically, this ratio goes to the GC results being a constant
(or slightly depending on a particular choice of the freeze-out
state). However,  in dynamical models this fall may be energy
dependent. In particular, in the expanding fireball
model~\cite{TNFNR04} the strangeness correlation volume was
associated with that of the initial thermalized state as the
Lorentz-contracted cylinder $V_c=\pi R^2 \cdot 2R/\gamma_{cm}$,
where $\gamma_{cm}$ is the $\gamma$-factor of colliding nuclei in the
 c.m. system.
 In this case, for high colliding
energies the $K^+/\pi^+$ excitation function falls down faster
than for a constant volume. Nevertheless, the observed  peaked
maximum in the excitation function is not reproduced within this
dynamical model (see also the comprehensive analysis
\cite{BGKMS04}). It is remarkable that such a behavior has really
been predicted in terms of a simple statistical model of the early
stage~\cite{GG99} as the onset  of the first order deconfinement
phase transition. It is of interest to note that the above
mentioned calculations, made in a similar scenario but in a more
elaborated model of an expanding fireball~\cite{TNFNR04} with
various equations of state, do not give rise to a peaked maximum
and calculated excitation functions are practically independent of
the equation of state. In other words, the condition of global
strangeness neutrality at the freeze-out washes out
particularities of strangeness evolution at the earlier stage.
\begin{figure}[hb]
\includegraphics[width=15cm]{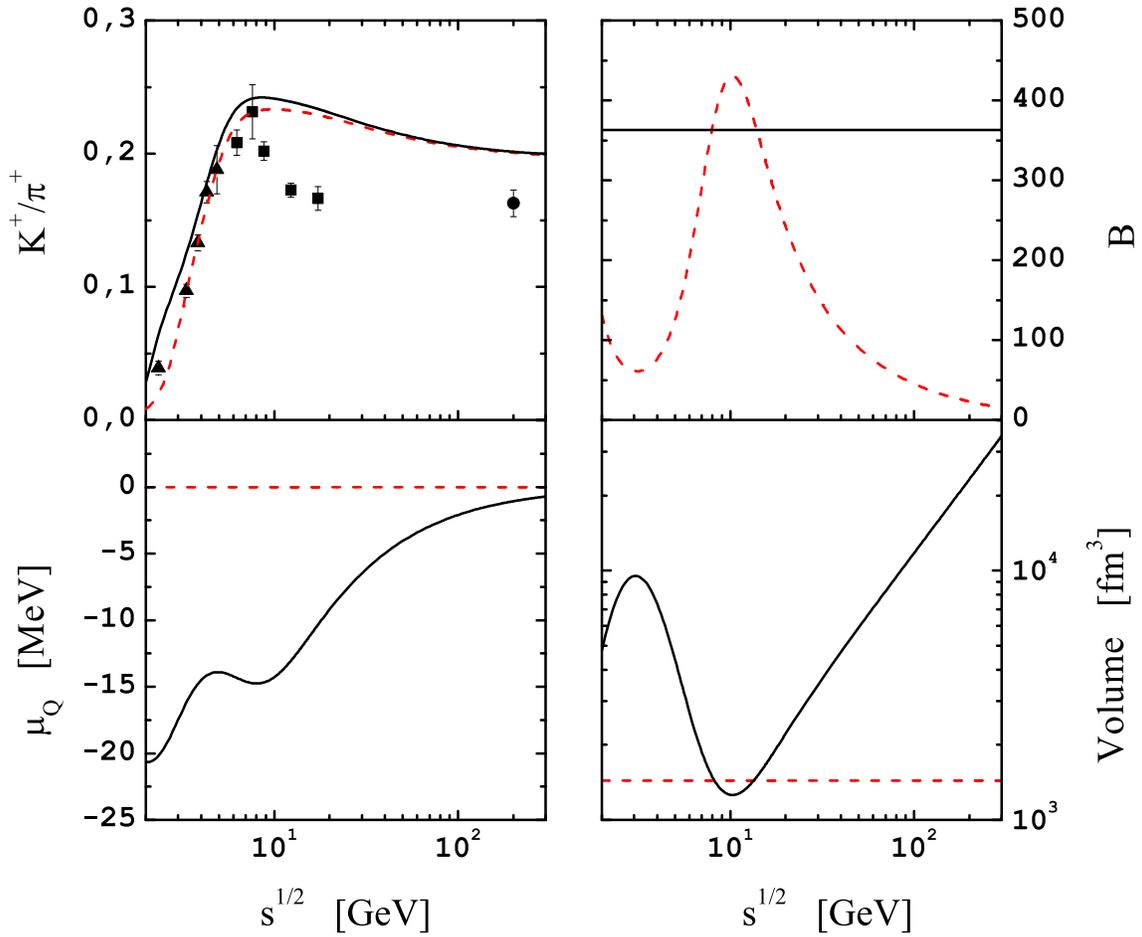}
  \caption{The $K^{+}/\pi^{+}$ excitation function and energy dependence of
  the total baryon charge $B$, electrical chemical potential
  $\mu_{Q}$ and volume (sphere radius) for the net strangeness of the
  system $S=0$. Lines are statistical model calculations for classical
  statistics and $s_{max}=1$ ({\it i.e.} multi-strange hyperons are
  ignored)  at two  assumptions~: {\em (i)} dashed 
  lines -- $\mu_{Q}=0$ and correlated  sphere radius $R=7$ fm (as in
  Fig.1); {\em (ii)} solid lines -- $\mu_{Q}$ and $R$  are found from the
additional conservation of baryon and electric charges of the
system, Eqs.(\ref{2f}) and (\ref{3f}).  Experimental points are the
same as in Fig.1.}
   \label{2}
\end{figure}

 Generally speaking, the introduction of two volumes, $V$ and $V_c$, into
 the statistical description based on the empirical freeze-out $T-\mu_B$ curve
 breaks thermodynamical consistency of the model. Indeed, in the presented
 calculations there is only $V$. On the other hand, the $T$ and
 $\mu_B$ values were  extracted from the particle ratios and the volume
 is not fixed.  To correct the description we introduce explicitly the
 conservation of  the total baryon $B$  and electric $Q$ charges
\begin{eqnarray}\label{2f}
  B &=& \sum_{\alpha}  \langle n_{\alpha}\rangle b_{\alpha} , \\
  Q &=& \sum_{\alpha}  \langle n_{\alpha}\rangle q_{\alpha}~.
  \label{3f}
\end{eqnarray}
Since $\mu_B$ is known from the $T-\mu_B$ curve, the first
equation (\ref{2f}) defines the volume of the system and the second
one allows one to find the electric charge potential $\mu_Q$. The
kaon excitation function recalculated for new obtained volume
$V=V_c$ is presented in Fig.2 together with some thermodynamic
quantities. These results are obtained for $B=$364 and $Q=$142 due to
5$\%$ centrality selection of data.

It is seen that the recalculated $K^{+}/\pi^{+}$ ratio does not
differ noticeably from that in the case of free correlation volume
$V_c$, except maybe the SIS energies due to non-vanishing
$\mu_{Q}$. The obtained electric chemical potential is not large
and it is getting even smaller with increasing the invariant
colliding energy
 $s^{1/2}$. However, the baryon conservation
is fulfilled only for the energy dependent volume. It is noteworthy that
the system volume behaves non-monotonically having a minimum near $s^{1/2}
\sim 10$ GeV. This finding may be an evidence of the softest point effect
at the passage  of the phase boundary. In addition, this behavior
roughly correlates with the freeze-out pion volume measured by HBT
interference experiments~\cite{TW02}.

Coming back to Fig.1, one can compare the $K^{+}/\pi^{+}$
excitation function to those calculated within the
Maxwell-Boltzmann ensemble but for different values of net
strangeness $S$ of the hadronic system. It is not of purely academic
interest. As was first noted in~\cite{SSG86}, in the Gibbs phase
coexistence of a baryon-rich quark-gluon plasma there is
strangeness separation in phases, the so-called {\em strangeness
distillation} effect: While the net strangeness of the system is
zero, the strangeness of its each component can differ from zero.
In particular, in the quark subsystem, being a small admixture to
the hadronic one, a number of strange quarks can be larger than
that of strange antiquarks ({\it i.e.} $S<0$) which should be
compensated by hadronic phase ({\it i.e.} $S>0$). A similar situation
may occur below 
the critical temperature for systems with crossover deconfinement
phase transition. The distillation effect is even stronger for
lower colliding energies, as  recently shown in~\cite{TNFNR04}. In
contrast to the first order phase transition, here a small quark
admixture can survive during the expansion stage until freeze-out.
This may give rise to an observable effect if particles from each
phase get on shell at freeze-out preserving their
strangeness. One should remember that recent lattice QCD
calculations with physical quark masses~\cite{ZK04} tell us that
at $\mu_B=360\pm 40$ MeV and $T=162\pm 2$ MeV there is the
critical end-point where the crossover changes to a phase transition
of the first order. It means that proceeding from high to lower
energies (see Fig.1), the hadronic subsystem  follows along the
curve for $K^+/\pi^+$ with $S>0$ strangeness components, then
stops near the end-point and jumps down on the curve with $S=0$.
The given parameters of the critical end-point roughly correspond
to the bombarding energy ~$\sim 30$ GeV~\cite{TNFNR04} and
correlate with the position of the maximum for the relative
strangeness yield in Fig.1. Therefore, in this scenario the
observed maximum is treated as a manifestation of the critical
end-point rather than a jump due to the latent heat at the first
order phase transition. A change of the  nature of phase transition  
at the critical end-point may  manifest itself in other
observables as well. Since at this change the  dominating
hadronic phase of the quark-hadron mixture is enriched with strange
antiquarks, effects should be more noticeable in the $K^+$ component.

\section{Conclusions}

The derivation of the Maxwell-Boltzmann partition function and
ensemble averages with the exact strangeness conservation has been
reconsidered in a general case of non-vanishing net strangeness. A
new method of recurrence relations was applied to solve the
partition function. This method has certain advantages as compared
to the numerical method used in~\cite{BGKMS04} for the canonical
description of baryon charge conservation of systems with the
total baryon charge up to about 100. The recurrence relation
technique can be successfully applied to larger systems, as it was
earlier demonstrated by a canonical treatment of statistical
multifragmentation of excited heavy nuclei~\cite{PTP00}.

Using the recurrence relations method we first generalize the
calculation scheme for the strangeness canonical ensemble to the
case of quantum ensemble. Taking into account quantum statistics,
the $K^+/\pi^+$ ratio turns out to be slightly lower  at collision
energies $\sqrt s \gsim 6$ GeV.

In the present approach the general behavior of the $K^+/\pi^+$
experimental excitation function  is reproduced roughly to the same
level of agreement as that in earlier statistical treatments.
Uncertainties related to a possible contribution of weak decays and
 energy dependence of dynamic correlation volume are pointed out.
 While the initial state of the equilibrium system is taken from the
 chemical freeze-out $T-\mu_B$ curve, an additional account of the
 baryonic and electric charge conservation does not change
 noticeably the particle ratio but provides the thermodynamic
 consistency of the strangeness canonical approach. The obtained
 unique system volume is energy dependent and exhibits clear minimum
 near $s^{1/2} \sim 10$ GeV,  which deserves further investigations.

 Nevertheless, the anomalous increase of this ratio near the
 bombarding energy of about 30 GeV, having been interpreted as a sign of
 deconfinement phase transition, is not reproduced by this model.
 A new scenario based on the strangeness distillation idea has been proposed
 where this anomalous strangeness enhancement is considered as a sign 
of the evolving system passing in vicinity of the critical end-point.
 Arguments in favor of this hypothesis are given. In this respect,
 multiplicity fluctuations of rare strange particles can be a measure of
 chemical equilibrium~\cite{JKRW02}, and therefore are of special
 interest. A more convincing test needs delicate dynamic calculations
 which are in  progress now.

{\bf Acknowledgements:}\\  We are grateful to B.Friman, Yu.Ivanov and
 K.Redlich for discussion  and valuable remarks.
 This work was supported in part by the Deutsche Forschungsgemeinschaft
(DFG project 436 RUS 113/558/0-2), the Russian Foundation for Basic
 Research (RFBR grants 03-02-04008) and  the Moldavian-U.S. Bilateral
 Grants Program (CRDF Project MP2-3045).

\appendix
\section{Recursive relations in classical statistics}
Let us prove the recurrence relations  (\ref{8a}) for the classical
perfect gas partition function with positive and negative components
of the net strangeness. It is easy to see that for the given species
$\alpha$ in a quantum state  $p$ the sum over occupation numbers
$\nu_{\alpha p}$ satisfies the following identity ~\cite{Par04}
\begin{eqnarray}\label{1c}
&& \sum\limits_{\nu_{\alpha p}=0}^{\infty}
\delta(\sum\limits_{\alpha^{\prime},p^{\prime}}
s_{\alpha^{\prime}}\nu_{\alpha^{\prime} p^{\prime}} - S) \
\nu_{\alpha p} \ \frac{1}{\nu_{\alpha p}!} \ e^{-\beta \nu_{\alpha
p} \varepsilon_{\alpha p}}= \nonumber \\
&& \;\;\;\;\;\;\;  = \ e^{-\beta \varepsilon_{\alpha p}}
 \sum\limits_{\nu_{\alpha p}=0}^{\infty}
 \delta(\sum\limits_{\alpha^{\prime},p^{\prime}}
s_{\alpha^{\prime}} \ \nu_{\alpha^{\prime} p^{\prime}} -
(S-s_{\alpha})) \frac{1}{\nu_{\alpha p}!} \ e^{-\beta \nu_{\alpha p}
 \varepsilon_{\alpha p}}~.
\end{eqnarray}
Taking into account the Kronecker symbol, Eq.(\ref{6a}) can be
rewritten as
\begin{eqnarray}\label{2c}
Z_{S_{\mp}}^{(\mp)} &=&
\frac{1}{S_{\mp}}\sum\limits_{\alpha^{\prime}\in
M_{\mp}}\sum_{p^{\prime}} |s_{\alpha^{\prime}}|
\sum\limits_{\{\nu_{\alpha p}\}_{\alpha\in M_{\mp}}}
\delta(\sum\limits_{\alpha\in M_{\mp}}\sum_{p} |s_{\alpha}|
\nu_{\alpha p} - S_{\mp})\times \nonumber \\  && \;\;\;\;\;\;
\times \ \nu_{\alpha^{\prime} p^{\prime}} \
\frac{1}{\prod\limits_{\alpha\in
M_{\mp}}\prod\limits_{p}\nu_{\alpha p}!} \
e^{-\beta\sum_{\alpha\in M_{\mp}}\sum_{p}\nu_{\alpha
p}\varepsilon_{\alpha p}}~.
\end{eqnarray}
 Using identity (\ref{1c}), the summation over $\nu_{\alpha p}$
gives
\begin{eqnarray}\label{3c}
Z_{S_{\mp}}^{(\mp)} &=&
\frac{1}{S_{\mp}}\sum\limits_{\alpha^{\prime}\in
M_{\mp}}\sum_{p^{\prime}} |s_{\alpha^{\prime}}| e^{-\beta
\varepsilon_{\alpha^{\prime} p^{\prime}}}
\sum\limits_{\{\nu_{\alpha p}\}_{\alpha\in M_{\mp}}}
\delta(\sum\limits_{\alpha\in M_{\mp}}\sum_{p} |s_{\alpha}|
\nu_{\alpha p} - (S_{\mp}-|s_{\alpha^{\prime}}|)) \times \nonumber\\
&& \;\;\;\;\;\;\;\;\;\;\;\;\;  \times \
\frac{1}{\prod\limits_{\alpha\in
M_{\mp}}\prod\limits_{p}\nu_{\alpha p}!} \
e^{-\beta\sum_{\alpha\in M_{\mp}}\sum_{p}\nu_{\alpha p}
\varepsilon_{\alpha p}}~.
\end{eqnarray}
By means of Eq.(\ref{1a}) the recurrence relations (\ref{3c}) can be
obtained directly~\cite{PTP00,Par04}
\begin{eqnarray}\label{4c}
Z_{S_{\mp}}^{(\mp)} &=& \frac{1}{S_{\mp}}\sum\limits_{\alpha\in
M_{\mp}}\sum_{p} |s_{\alpha}| \ e^{-\beta\varepsilon_{\alpha p}}
Z_{S_{\mp}-|s_{\alpha}|}^{(\mp)}~.
\end{eqnarray}
Thus, the formulae (\ref{8a}) has been proved.

Mean occupation numbers are evidently given as
\begin{eqnarray}\label{5c} \langle
  \nu_{\alpha^{\prime}p^{\prime}}\rangle_{S}  &=&
\frac{1}{Z_S}\sum\limits_{\{\nu_{\alpha p}\}}
\delta(\sum\limits_{\alpha,p} s_{\alpha} \nu_{\alpha p} - S) \
\nu_{\alpha^{\prime} p^{\prime}} \ \frac{1}{\prod\limits_{\alpha,p}\
\nu_{\alpha p}!} \ e^{-\beta\sum_{\alpha,p}\nu_{\alpha
p}\varepsilon_{\alpha p}}~.
\end{eqnarray}
To sum here over $\nu_{\alpha^{\prime} p^{\prime}}$ the identity
(\ref{1c}) should be used~:
\begin{eqnarray}\label{6c} \langle
  \nu_{\alpha^{\prime}p^{\prime}}\rangle_{S}  &=&
e^{-\beta\varepsilon_{\alpha^{\prime} p^{\prime}}}
\frac{1}{Z_S}\sum\limits_{\{\nu_{\alpha p}\}}
\delta(\sum\limits_{\alpha,p} s_{\alpha} \nu_{\alpha p} -
(S-s_{\alpha^{\prime}})) \ \frac{1}{\prod\limits_{\alpha,p}\
\nu_{\alpha p}!} \ e^{-\beta\sum_{\alpha,p}\nu_{\alpha
p}\varepsilon_{\alpha p}} =  \\ \label{7c}&& \;\;\;\;\;\;\;\;\;\;\;
= \ e^{-\beta \varepsilon_{\alpha' p'}} \
\frac{Z_{S-s_{\alpha'}}}{Z_{S}}~,
\end{eqnarray}
where Eq.(\ref{1a}) has been used  for getting the latter equality.

\section{Recursive relations in quantum statistics}
Here we will prove the recurrence relations (\ref{3b}) for the
quantum partition functions with positive and negative components of
net strangeness. It is readily seen that the following identity for
occupation number summing is valid~\cite{Par04}
\begin{eqnarray}\label{1d}
&& \sum_{\nu_{\alpha p}=0}^{K_{\alpha}} \nu_{\alpha p} \ e^{-\beta
\nu_{\alpha p} \varepsilon_{\alpha p}}
\delta(\sum_{\alpha',p'}|s_{\alpha'}| \ \nu_{\alpha' p'}-S_{\pm})= \nonumber \\
 && \;\;\;\;\; = \ \sum_{l=1}^{[S_{\pm}/|s_{\alpha}|]} y_{\alpha l} \
e^{-\beta l  \varepsilon_{\alpha p}} \sum_{\nu_{\alpha
p}=0}^{\min([S_{\pm}/|s_{\alpha}|]-l,K_{\alpha})} e^{-\beta
\nu_{\alpha p} \varepsilon_{\alpha p}} \ \delta(\sum_{\alpha',p'}|
s_{\alpha'}| \ \nu_{\alpha' p'}-(S_{\pm}-l|s_{\alpha}|))~.
\end{eqnarray}
Using the Kronecker symbol, we can rewrite Eq.(\ref{1bb})~:
\begin{eqnarray}\label{2d}
Z_{S_{\mp}}^{(\mp)} &=& \frac{1}{S_{\mp}}\sum\limits_{\alpha'\in
M_{\mp}}\sum_{p'} |s_{\alpha'}| \sum\limits_{\{\nu_{\alpha
p}\}_{\alpha\in M_{\mp}}} \delta(\sum\limits_{\alpha\in
M_{\mp}}\sum_{p} |s_{\alpha}| \ \nu_{\alpha p} - S_{\mp}) \
\nu_{\alpha' p'}  \ e^{-\beta\sum_{\alpha\in
M_{\mp}}\sum_{p}\nu_{\alpha p} \ \varepsilon_{\alpha p}}~.
\end{eqnarray}
Applying identity (\ref{1d}) for summing over $\nu_{\alpha p}$, we
have
\begin{eqnarray}\label{3d}
Z_{S_{\mp}}^{(\mp)} &=& \frac{1}{S_{\mp}}\sum\limits_{\alpha'\in
M_{\mp}}\sum_{p'} |s_{\alpha'}|
\sum_{l=1}^{[S_{\mp}/|s_{\alpha'}|]} y_{\alpha' l} e^{-\beta l
\varepsilon_{\alpha' p'}} \\ \nonumber &\times&
\sum\limits_{\{\nu_{\alpha p}\}_{\alpha\in M_{\mp}}}
\delta(\sum\limits_{\alpha\in M_{\mp}}\sum_{p} |s_{\alpha}| \
\nu_{\alpha p} - (S_{\mp}-l |s_{\alpha'}|))  \
e^{-\beta\sum_{\alpha\in M_{\mp}}\sum_{p}\nu_{\alpha
p}\varepsilon_{\alpha p}}~.
\end{eqnarray}
The use of (\ref{1b}) in Eq.(\ref{3d}) immediately gets the
recurrence relations~\cite{PTP00,Par04}
\begin{eqnarray}\label{4d}
Z_{S_{\mp}}^{(\mp)} &=& \frac{1}{S_{\mp}}\sum\limits_{\alpha\in
M_{\mp}}\sum_{p} |s_{\alpha}| \sum_{l=1}^{[S_{\mp}/|s_{\alpha}|]}
y_{\alpha l} \ e^{-\beta l \varepsilon_{\alpha p}} \ Z_{S_{\mp}-l
|s_{\alpha}|}^{(\mp)}~,
\end{eqnarray}
which completes the derivation of Eq.(\ref{3b}).

The partition function with zero strangeness component has the form
\begin{equation}\label{5d}
  Z^{(0)}=\sum_{\{\nu_{\alpha p}\}_{\alpha\in M_{0}}}
  e^{-\beta\sum_{\alpha\in M_{0}}\sum_{p}
  \nu_{\alpha p} \varepsilon_{\alpha p}}.
\end{equation}
This partition function (\ref{5d}) can be represented as the product
of sums for every species  $\alpha$ and then, accounting for the
Kronecker symbols, be reduced to the equation
\begin{equation}\label{6d}
Z^{(0)}=\prod_{\alpha\in
M_{0}}\sum_{n_{\alpha}=0}^{\infty}\sum_{\{\nu_{\alpha
p}\}'}\delta(\sum_{p}\nu_{\alpha p}-n_{\alpha}) \ e^{-\beta\sum_{p}
  \nu_{\alpha p} \varepsilon_{\alpha p}}~,
\end{equation}
where  $\{\nu_{\alpha p}\}'$ is the set of occupation numbers for a
fixed species $\alpha$. Formally, sums over occupation numbers
$\{\nu_{\alpha p}\}'$ in Eq.(\ref{6d}) are  quantum single-particle
partition function $n_{\alpha}$ of identical particles in the
canonical ensemble (see \cite{PTP00,Par04}). So, Eq.(\ref{6d}) can
be represented as follows
\begin{equation}\label{7d}
Z^{(0)}=\prod_{\alpha\in
M_{0}}\sum_{n_{\alpha}=0}^{\infty}\sum_{\{N_{\alpha
l}\}'}\delta(\sum_{l=1}^{n_{\alpha}}l N_{\alpha l}-n_{\alpha})
\prod_{l=1}^{n_{\alpha}} \frac{(Z_{\alpha}^{l})^{N_{\alpha l
}}}{l^{N_{\alpha l}}N_{\alpha l}!}~.
\end{equation}
After simple transformations we finally get
\begin{equation}\label{8d}
Z^{(0)}=\prod_{\alpha\in M_{0}}\prod_{l=1}^{\infty}e^{l^{-1} \
Z_{\alpha}^{l}}~.
\end{equation}

Mean occupation number are given by the following expression
\begin{eqnarray}\label{9d} \langle \nu_{\alpha p}\rangle_{S}  &=&
\frac{1}{Z_S}\sum\limits_{\{\nu_{\alpha p}\}}
\delta(\sum\limits_{\alpha,p} s_{\alpha} \nu_{\alpha p} - S) \
\nu_{\alpha p}  \ e^{-\beta\sum_{\alpha,p}\nu_{\alpha p} \
\varepsilon_{\alpha p}} =-\frac{1}{\beta}\frac{\partial\ln
Z_{S}}{\partial\varepsilon_{\alpha p}}~.
\end{eqnarray}

With (\ref{4a}), from here we get for mean occupation numbers with
$\alpha\in M_{\mp}$
\begin{eqnarray}\label{10d}
\langle \nu_{\alpha p}\rangle_{S}&=& \frac{Z^{(0)}}{Z_{S}}
\sum_{S_{+}=0}^{\infty}\sum_{S_{-}=0}^{\infty}
\delta(-S_{-}+S_{+}-S) \ Z_{S_{\pm}}^{(\pm)} \times \nonumber  \\
&\times & \sum\limits_{\{\nu_{\alpha p}\}_{\alpha\in M_{\mp}}}
\delta(\sum\limits_{\alpha\in M_{\mp}}\sum_{p} |s_{\alpha}| \
\nu_{\alpha p} - S_{\mp}) \ \nu_{\alpha p}  \
e^{-\beta\sum_{\alpha\in M_{\mp}}\sum_{p}\nu_{\alpha
p}\varepsilon_{\alpha p}}, \;\;\;\;\;\;\; \alpha\in M_{\mp}~.
\end{eqnarray}
Using identity (\ref{1d}) and following the derivation of Eqs.(\ref{2d})
-- (\ref{4d}) from Eq.(\ref{10d}), we arrive at Eq.(\ref{7b}) for
the mean occupation numbers.

In the case of  $\alpha\in M_{0}$, the mean occupation numbers are
given by the equation
\begin{equation}\label{11d}
\langle \nu_{\alpha p}\rangle_{S}  =
\frac{1}{Z^{(0)}}\sum\limits_{\{\nu_{\alpha p}\}_{\alpha\in
M_{0}}} \nu_{\alpha p} \ e^{-\beta\sum_{\alpha\in
M_{0}}\sum_{p}\nu_{\alpha p}\varepsilon_{\alpha
p}}=-\frac{1}{\beta}\frac{\partial\ln
Z^{(0)}}{\partial\varepsilon_{\alpha p}}, \;\;\;\;\;\;\; \alpha\in
M_{0}~.
\end{equation}
If the partition function (\ref{8d}) for the zero strangeness
component is substituted into differential relation (\ref{11d}),
then we arrive at Eq.(\ref{8b}) for mean occupation numbers with
$\alpha\in M_{0}$.


\begin{thebibliography}{99}

\bibitem{BRMS03} Braun-Munzinger P, Redlich K and Stachel J
in {\em Quark Gluon Plasma 3} eds. Hwa R C and Wang  X N,
World Scientific, Singapore p.491; [nucl-th/0304013]
%
\bibitem{RT80} Redlich K and  Turko B  1980 {\em Z. Phys.} {\bf B 97} 279;
Turko B 1981 {\em Phys. Lett.} {\bf B 104} 153
%
\bibitem{HR85} Hagedorn R and Redlich R  1985 {\em Z. Phys.} {\bf C 27} 541;
Rafelski J and Danosh D 1980 {\em Phys. Lett.} {\bf B 97} 279; Turko B and
Rafelski J 2001{\em  Eur. Phys. J.} {\bf C 18} 587
%
\bibitem{RKT03}
 Redlich K, Karsch F and Tounsi A  2002
 In {\em Physical and Mathematical Aspects of Symmetries}
 Paris p.139;  [hep-ph/0302245]
%
\bibitem{KB02-1} Ker\"{a}nen A and Becattini F  2002
{\em J.Phys.} {\bf G28} 2041; [nucl-th/0112045]
%
\bibitem{CR99} Cleymans J, Redlich R and Suhonen E  1991 {\em Z. Rev.}
{\bf C 51} 137;  Cleymans J, Oeschler H and Redlich K  2001
  {\em  Phys. Lett.} {\bf B 485} 27
%
\bibitem{HRT00-1} Hamieh S, Redlich K and Taunsi A 2000
 {\em Phys. Lett.} {\bf B 486}  61; 2001 {\em J. Phys.} {\bf G 27}  413
%
\bibitem{R58} Reordan J 1958 {\em An Introduction to Combinatorial Analysis}
Wiley, New York
%
\bibitem{PTP00} Parvan A S, Toneev V D and P{\l}oszajczak M 2000
{\em Nucl. Phys.} A {\bf 676} 409

\bibitem{Par04} Parvan A S 2004 {\em Theor. and Math. Phys.} {\bf 140}
 977
%
\bibitem{Becat} Ker\"{a}nen A and Becattini F   2002
{\em Phys. Rev.} C {\bf 65} 044901; [nucl-th/0112021]
%
\bibitem{CSW92} Cleymans J Suhonen E and Weber G M 1992 Z. Phys. C
  {\bf 53} 485  
%
\bibitem{Mek} Chase K C and Mekjian A Z 1994 {\em Phys. Rev.} C {\bf 50}
 2078;  1995 {\em Phys. Rev.} C {\bf 52} R2339
%
\bibitem{PR03} Pratt S and Ruppert J 2003 {\em Phys. Rev.} C {\bf 68} 024904
%
\bibitem{Hagiwara} Hagiwara K {\em et al} 2002 {\em Phys. Rev.} D {\bf
    66}  010001-1
%
\bibitem{Andronic} Andronic A and Braun-Munzinger P 2004 hep-ph/0402291
%
\bibitem{Brat1} Bratkovskaya E L, Bleicher M, Reiter M, Soff S,
St\"{o}cker H, van~Leeuwen M, Bass S A and  Cassing W 2004 {\em Phys.Rev.} C
{\bf 69} 054907
%
\bibitem{BGKMS04} Becattini F, Gazdzicki M, Keranen A, Manninen J and
Stock R 2004
{\em Phys.Rev.} C {\bf 69} 024905; [hep-ph/0310049]
%
\bibitem{B-NA49} Botje M, for the NA49 Collaboration 2004
nucl-ex/0407004
%
\bibitem{TR02} Taunsi A and Redlich K 2002
 {\em J. Phys.} {\bf G 28} 2095
%
\bibitem{TNFNR04} Toneev V D, Nikonov E G, Friman B, N\"orenberg W
and Redlich K 2004
{\em Eur. Phys. J.} {\bf C32}  399; [hep-ph/0308088]
%
\bibitem{GG99} Gazdzicky M and Gorenstein M I 1999 {\em Acta
    Phys. Pol.}
{\bf B 30} 2705; [hep-ph/9803462]
%
\bibitem{TW02} Tomasik B and Wiedemann U A
in {\em Quark Gluon Plasma 3}, eds.  Hwa R C and  Wang X N,
World Scientific, Singapore; [nucl-th/0304013]
%
\bibitem{SSG86} Subramanian P R, St\"{o}cker H and Greiner W 1986 {\em
 Phys. Lett.} {\bf 173} 468; Greiner C, Koch P and St\"{o}cker H 1987
{\em Phys. Rev. Lett.} {\bf
58} 1825; 1988 {\em Nucl. Phys.} {\bf A 479} 295c
%
\bibitem{ZK04} Fodor Z and Katz S D  2004
{\em JHEP} {\bf 404} 50; [hep-lat/0402006]
%
\bibitem{JKRW02} Jeon S, Koch V, Redlich K and Wang X N 2002
 {\em Nucl.Phys.} {\bf A697}  546; [nucl-th/0105035]

\end{thebibliography}
\end{document}